\documentstyle[floats,aps,aipbook,psfig]{revtex}

\def\aboutless{\ifmmode {\mathbin{\lower 3pt\hbox
    {$\rlap{\raise 5pt\hbox{$\char'074$}}\mathchar"7218$}}}   
    \else {${\mathbin{\lower 3pt\hbox
    {$\rlap{\raise 5pt\hbox{$\char'074$}}\mathchar"7218$}}}   
    $\thinspace}\fi}
\def\aboutmore{\ifmmode {\mathbin{\lower 3pt\hbox
    {$\rlap{\raise 5pt\hbox{$\char'076$}}\mathchar"7218$}}}   
    \else {${\mathbin{\lower 3pt\hbox
    {$\rlap{\raise 5pt\hbox{$\char'076$}}\mathchar"7218$}}}   
     $\thinspace}\fi}

\begin{document}

\title{Compton scattering effects in the\\
	spectra of soft gamma-ray repeaters}

\author{M. Coleman Miller\thanks{{\it Compton} GRO Fellow}
        and Tomasz Bulik\thanks{Also at: N. Copernicus Astronomical
Center, Bartycka 18, 00-716, Warsaw, Poland.}}

\address{Department of Astronomy and Astrophysics \\
	University of Chicago, Chicago, Illinois 60637}

\maketitle
\begin{abstract}

The association of all three soft gamma-ray repeaters (SGRs) with supernova
remnants has made possible estimates of the distance to and luminosity of
the sources of SGRs, which have provided a starting point for detailed
modeling.
One of the most popular classes of models involves strongly magnetized
neutron stars, with surface dipole fields $B\sim 10^{14}-10^{15}$ G.
In these ``magnetar" models, many otherwise negligible processes
can play an important role.  Here
we consider the spectral effects of strong-field modifications to Compton
scattering, in particular those related to the contribution of vacuum
polarization to the dielectric tensor.  Vacuum polarization introduces
a density-dependent photon frequency, called the second vacuum frequency,
at which the normal modes of polarization become nonorthogonal and the
mean free path of photons decreases sharply.  Monte Carlo simulations
of photon propagation through a magnetized plasma show that this effect
leads, under a wide range of physical conditions, to a broad
absorption-like feature in the energy range $\sim$5 keV---40 keV.

\end{abstract}

\section*{Introduction}

   The first soft gamma-ray repeater (SGR) to be discovered,
SGR 0526-66 \cite{Maz79,Maz81},
was shown soon after its discovery to be positionally coincident with
the N49 supernova
remnant in the Large Magellanic Cloud \cite{E80}.
More recently, the other two
SGRs (SGR 1806-20 and SGR 1900+14) have also been associated with
supernova remnants \cite{Kou93,Kou94,H94,KF93,Kul94}.  It is now
generally accepted that SGRs originate from young neutron stars.

   For a variety of reasons, summarized in \cite{TD95}, a popular class
of models for SGRs assumes that these neutron stars have very strong
surface magnetic fields, on the order of $10^{14}-10^{15}\,$G.
These models
are extremely complicated, and many of their properties depend on microscopic
physical processes that are important only in very strong magnetic fields.
Here we
concentrate on the effects of vacuum polarization and, in particular,
on the spectral effects of the second vacuum frequency.

   In strong magnetic fields $B\aboutmore B_c$, where $B_c=4.414\times
10^{13}\,$G is the magnetic field at which the electron cyclotron
energy equals the electron rest mass energy, vacuum polarization effects
due to virtual $e^+e^-$ pairs can become significant (see, {\it e.g.},
\cite{M92} for a discussion).  In fact, for strong enough fields
the vacuum contribution to the dielectric tensor can exceed the plasma
contribution \cite{GPS78,PS79}.
This affects the photon normal modes and cross sections.
Well below the electron cyclotron frequency $\omega_B\equiv eB/m_ec$, the
normal
modes for photons in a very strong magnetic field are nearly linearly
polarized over
most photon frequencies and incident angles.

   At photon frequencies $\omega\ll\omega_B$, the scattering
cross section depends strongly on the polarization of the photon.  If the
electric field vector of the photon is in the plane formed by the
magnetic field and the photon propagation direction $\hat{k}$, the photon
is in the
parallel mode and the scattering cross section $\sigma_\parallel\sim
\sigma_T$, where $\sigma_T=6.65\times 10^{-25}\,{\rm cm}^2$ is the Thomson
cross section.  However, if the electric field vector of the photon is
perpendicular
to the ${\vec B}-{\hat k}$ plane, the photon is in the perpendicular mode and
$\sigma_\perp\sim(\omega/\omega_B)^2\sigma_T\ll \sigma_T$.  This reduction
in the cross section is understandable, because in a strong magnetic field
it is easier to oscillate an electron along the field than to oscillate
it across the field.

There is, however, a small range of photon
frequencies in which the vacuum and plasma contributions to the dielectric
tensor are comparable to each other and the normal modes become strongly
nonorthogonal over a broad
range of angles.  This frequency, called the second vacuum frequency (the
first vacuum frequency is near the cyclotron frequency), depends on the
magnetic field and density: $\omega_{c2}\sim n^{1/2}B^{-1}$ for $B\ll B_c$
and $\omega_{c2}\sim n^{1/2}B^{-1/2}$ for $B\gg B_c$.  When $\omega
\approx \omega_{c2}$, $\sigma_\parallel
\approx \sigma_\perp\approx\sigma_T$. Figure~1 shows the dependence of
cross section on photon energy for both polarization modes for
$\rho=100$ g cm$^{-3}$ and $\rho=1000$ g cm$^{-3}$ when $B=10\,B_c$.
This figure shows that at any particular
density the cross section is unusually high for
a narrow range of frequencies.
Conversely, if we assume that most of the flux is
transported in the lower cross section mode, for any photon frequency there
is a density at which the cross section is high.  Thus, the optical depth
for a photon is greater than it would be in the absence of the enhanced
cross section at $\omega_{c2}$.  Because the frequency of the resonance varies
with density, in an atmosphere of varying density the vacuum resonance has
an effect on the spectrum in a wide range of photon energies.

\begin{figure}
\vspace{-1cm}
\centerline{\psfig{file=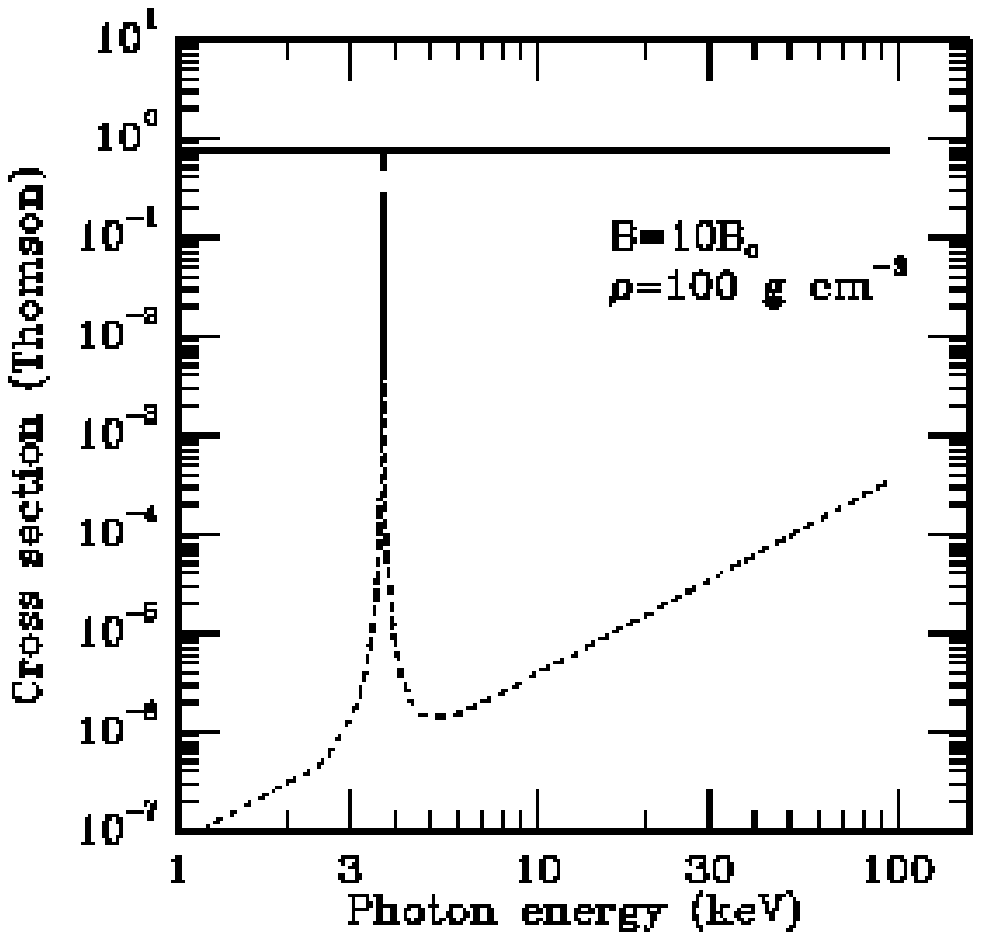,width=5.5cm,angle=0}\hfil
 \psfig{file=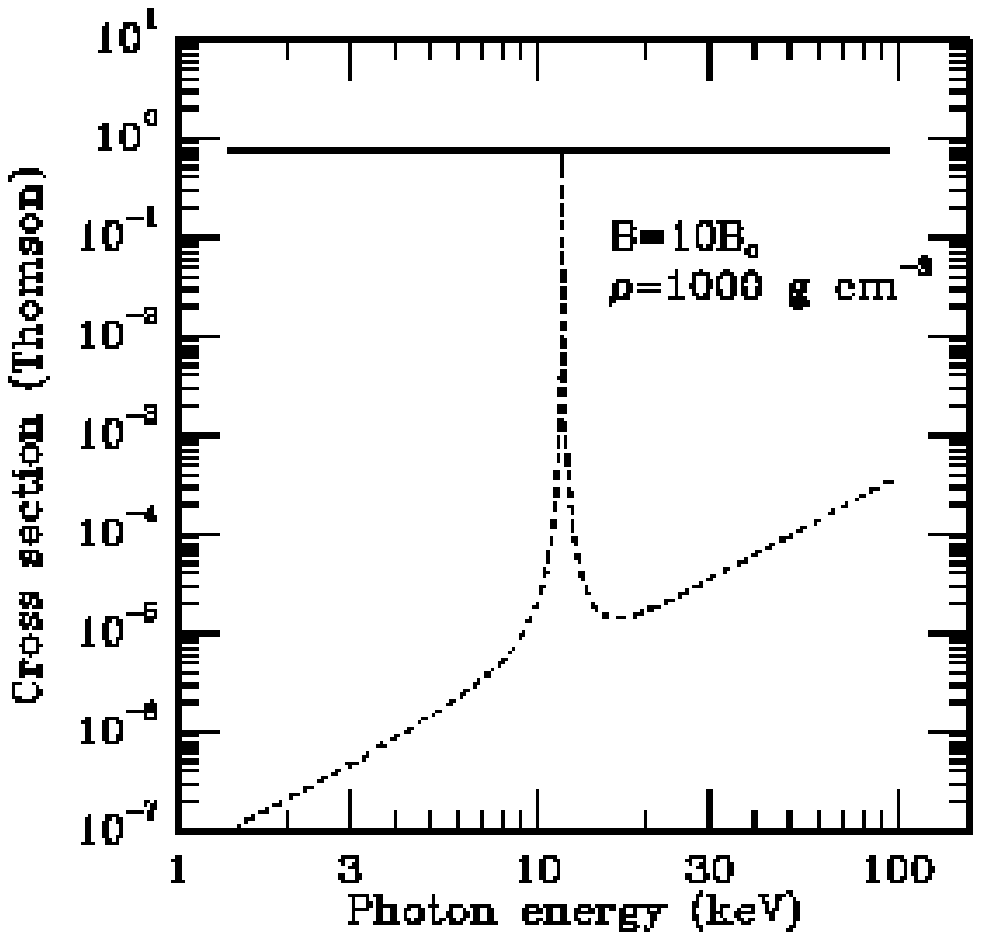,width=5.5cm,angle=0}}

\medskip
\caption{Cross section in units of the Thomson cross section versus
photon energy for the perpendicular (dotted line) and parallel (solid line)
polarization modes, for two different densities.  The plasma is assumed to
consist of fully ionized hydrogen, and the photon propagation
angle is $\theta=60^\circ$.}
\end{figure}

\section*{Energy range of the vacuum feature}

As described in, e.g., \cite{M92}, the scattering cross section depends on
the parameter
\begin{equation}
b
\approx{\sin^2\theta\over{2\cos\theta}}{\omega_B\over\omega}
\left[1-{\alpha\over{2\pi}}\left(\omega\over\omega_p\right)^2
(\eta_1-\eta_2)\right]\; ,
\end{equation}
where $\theta$ is the angle of propagation with respect to the magnetic
field, $\alpha$ is the fine structure constant, $\omega_p$ is the plasma
frequency, and $\eta_1$ and $\eta_2$ are functions of the magnetic
field related to the indices of
refraction of the two normal modes \cite{TE74}.  When $|b|\gg 1$ the
scattering
is non-resonant and the perpendicular mode has $\sigma\sim (\omega/
\omega_B)^2\sigma_T$.  When $|b|\ll 1$ the scattering is resonant and the
cross section in either mode is $\sigma\sim \sigma_T\sin^2\theta$.
The second vacuum frequency is
\begin{eqnarray}
\omega_{c2} &=& \omega_p \left(15\pi/\alpha\right)^{1/2}
\left(B_c/B\right)\qquad \ \ {\rm for\ }B\ll B_c\ ,\ {\rm and}\\
\omega_{c2} &=& \omega_p\left(3\pi/\alpha\right)^{1/2}
\left(B_c/B\right)^{1/2}\qquad {\rm for\ }B\gg B_c\; .
\end{eqnarray}
Because
$\omega_{c2}\sim n^{1/2}$, as radiation propagates from regions of high
density to regions of low density the frequency of the vacuum resonance
changes. In effect, the resonance acts as a sliding high opacity barrier.
Qualitatively, we expect that the resonance will lower the
flux at a given frequency if a) the optical depth through the resonance
is greater than unity, and b) the optical depth to the surface without the
resonance is less than unity.  These requirements give us, respectively,
the minimum and maximum photon energies at which the vacuum resonance
strongly affects the spectrum.

   As we show in \cite{BM96}, for $B\ll B_c$ and an atmosphere of
scale height $\ell$,
this simple analytical treatment gives
minimum and maximum photon energies
\begin{eqnarray}
\hbar\omega_{\rm min}&\approx& 6\left(B/B_c\right)^{-1/3}\left(\ell
/{\rm 10\ cm}\right)^{-1/3}\,
{\rm keV}\; ,\\
\hbar\omega_{\rm max}&\approx& 30\left(\ell/{\rm 10\ cm}\right)^{-1/4}
(\sin\theta)^{-1}\,{\rm keV}\; ;
\end{eqnarray}
if instead $B\gg B_c$,
\begin{eqnarray}
\hbar\omega_{\rm min}&\approx& 3\left(\ell/{\rm 10\ cm}\right)^{-1/3}
\, {\rm keV}\; ,\\
\hbar\omega_{\rm max}&\approx& 20\left(\ell/{\rm 10\ cm}\right)^{-1/4}
\left(B/B_c\right)^{1/4}(\sin\theta)^{-1}\,{\rm keV}\; .
\end{eqnarray}
The scalings with magnetic field and scale height are confirmed by
our numerical results (see \cite{BM96} and the following section):
typically $\hbar\omega_{\rm min}\approx 5$ keV and $\hbar\omega_{\rm max}
\approx 40$ keV, if the radiation is produced deep in the atmosphere.

\section*{Model spectra}

   To synthesize spectra, we use a Monte Carlo code and propagate
100,000 photons through an atmosphere that is exponential in density.
The input spectrum is a 10 keV blackbody, and for the spectra shown
here the magnetic field is $B=10\,B_c$.  In Figure~2 we show two of our
model spectra, one with a scale height $l=10$ cm and the other with
$l=1000$ cm.  We find (see also \cite{BM96}) that below $\sim$5 keV the
spectral shape is dominated by free-free absorption and induced scattering,
whereas above $\sim$5 keV direct Compton scattering is most important.
As expected, for photon energies between $\sim\hbar\omega_{\rm min}$ and
$\hbar\omega_{\rm max}$ the extra opacity created by the vacuum resonance
causes a deficit in the spectrum, and the upper range of this deficit decreases
with increasing scale height, as predicted by the analytical model above.
For $l=10$ cm an absorption-like feature is seen, whereas for $l=1000$ cm
the feature would be evident as a low-energy falloff.  In principle, a
transition from a falloff to an absorption-like feature would be expected
as the source goes from its bursting phase into an afterglow, and observation
of such a spectral transition would be evidence for very strong magnetic
fields.
Note also that the high-energy continuum is steeper than the input blackbody,
because the mean free path of photons to scattering goes as $\omega^{-2}$,
which shifts the distribution of escaping photons to lower energies.

   It is essential to include the effects of mode switching in these
simulations.  At temperatures of $\sim$10 keV the energy change
of a photon in a single scattering is much greater than the width of the
vacuum resonance.  Thus, in the $>$50\% of scatterings in the resonance
when the photon switches modes, the photon will be Comptonized
out of the resonance and may require several thousand scatterings
to return to the low cross section mode, during which its fractional
energy change can be of order unity.  However, the mean free path in the
parallel mode is small enough that the net distance traveled in the parallel
mode is negligible.

\begin{figure}
\vspace{-1cm}
\centerline{\psfig{file=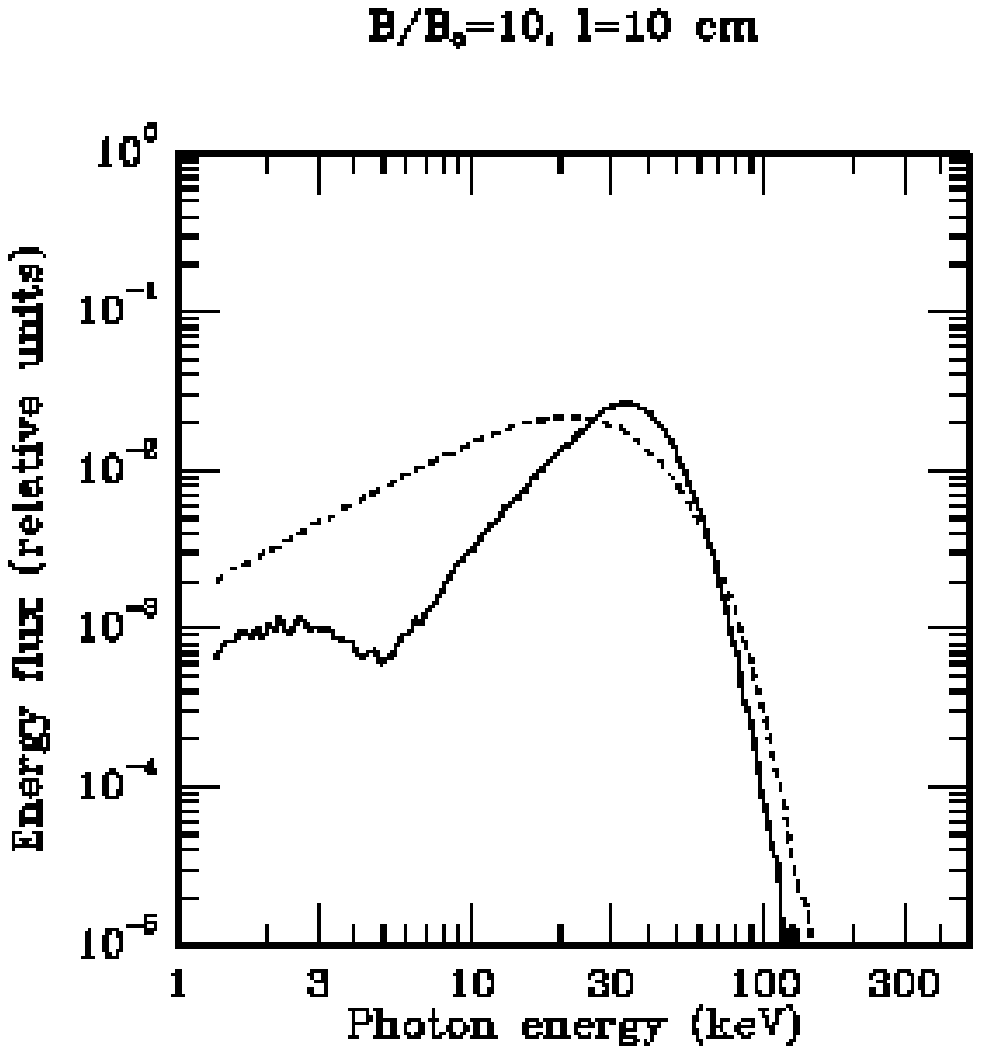,width=5.5cm,angle=0}\hfil
 \psfig{file=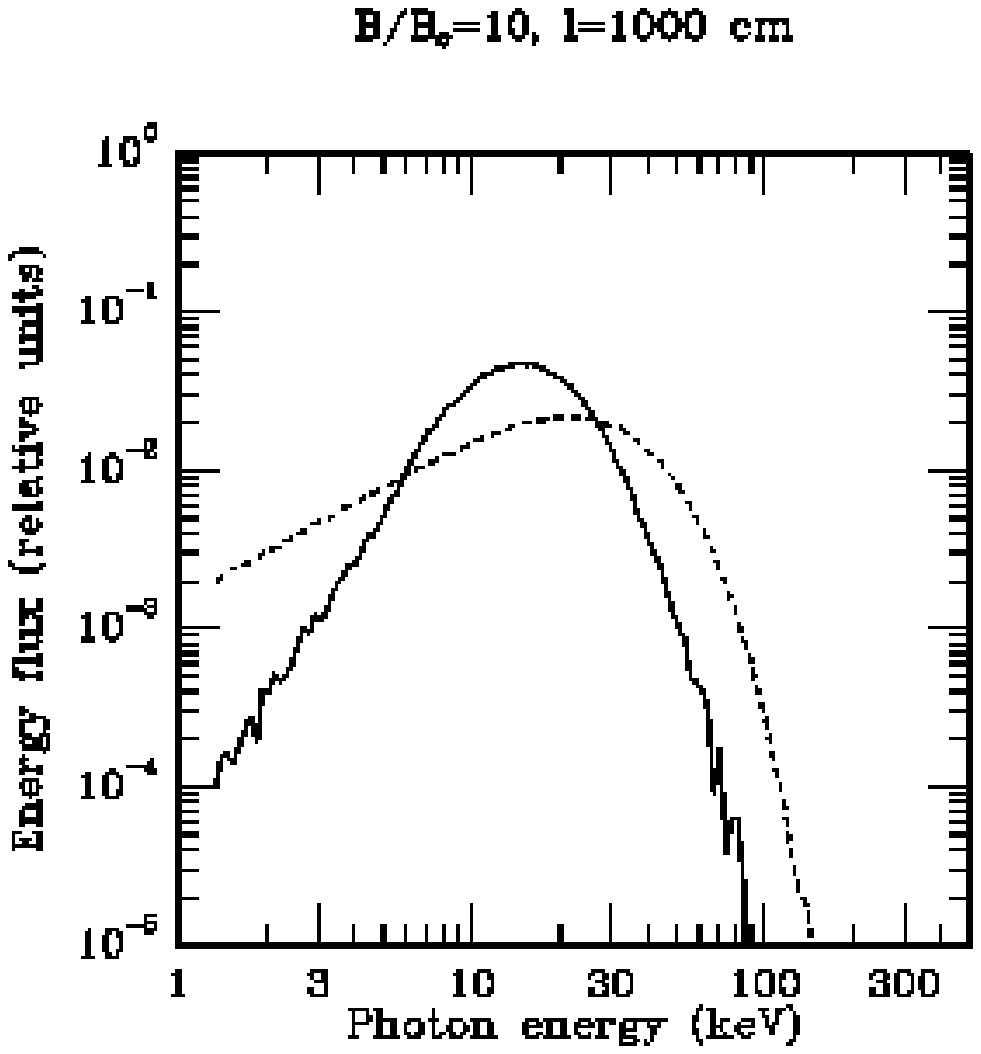,width=5.5cm,angle=0}}
\medskip
\caption{(left panel) Model spectrum seen at stellar surface (solid line)
for $B=10\,B_c$ and a scale height of $l=10$ cm.  Input spectrum (dotted
line) is a 10 keV blackbody.  (right panel) Same as left panel, except
that $l=1000$ cm.}
\end{figure}

\section*{Conclusions}

   For magnetic fields and plasma densities
like those expected in SGRs, photon scattering is strongly
affected by the vacuum resonance.  We find that the increased opacity
caused by this resonance typically creates a falloff or absorption-like
feature in the spectrum at $\sim 5$ keV to 40 keV.  The spectral energy and
shape of the feature depend on quantities such as the magnetic field
and scale height of the atmosphere.  Since scattering opacity is likely to
dominate over absorption opacity in the high-temperature environments of
SGRs, the effects of the vacuum resonance must be included in any detailed
calculations of radiative transfer through strongly magnetized atmospheres.

\end{document}